\documentclass[aps,pra,showpacs,reprint,floatfix,nobibnotes,superscriptaddress]{revtex4-1}
\usepackage{graphicx}
\graphicspath{{./figures_arXiv/}}
\usepackage{tabularx}

\usepackage{color}

\usepackage{braket}

\usepackage{amssymb}
\usepackage{amsthm}
\usepackage{amsmath}
\usepackage{upgreek}

\usepackage{units}
\usepackage{bpchem}
\usepackage[version=3]{mhchem}

\begin{document}

\title{X-Ray Studies of Nanoporous Gold: Powder Diffraction by Large Crystals with Small Holes}

\author{Matthias Graf}
\affiliation{Institute of Materials Physics and Technology, Hamburg University of Technology, Hamburg, Germany}

\author{Bao-Nam Dinh Ng{\^o}}
\affiliation{Institute of Materials Research, Materials Mechanics, Helmholtz-Zentrum Geesthacht, Geesthacht, Germany}

\author{J\"org Weissm\"uller}
\affiliation{Institute of Materials Physics and Technology, Hamburg University of Technology, Hamburg, Germany}
\affiliation{Institute of Materials Research, Materials Mechanics, Helmholtz-Zentrum Geesthacht, Geesthacht, Germany}

\author{J\"urgen Markmann}
\email{Corresponding author, email: juergen.markmann@hzg.de}
\affiliation{Institute of Materials Research, Materials Mechanics, Helmholtz-Zentrum Geesthacht, Geesthacht, Germany}
\affiliation{Institute of Materials Physics and Technology, Hamburg University of Technology, Hamburg, Germany}

\begin{abstract}
X-ray diffraction studies of nanoporous gold face the poorly understood diffraction scenario where large coherent crystals are riddled with nanoscale holes. Theoretical considerations derived in this study show that the ligament size of the porous network influences the scattering despite being quasi single crystalline. Virtual diffraction of artificially generated samples confirms the results but also shows a loss of long-range coherency and the appearance of microstrain due to thermal relaxation. Subsequently, a large set of laboratory X-ray investigations of nanoporous gold fabricated by different approaches and synthesis parameters reveal a clear correlation between ligament size and size of the coherent scattering domains as well as extremely high microstrains in samples with ligament sizes below \unit[10]{nm}.
\end{abstract}


\maketitle

\section{Introduction}

Nanoscale metal networks made by dealloying are of current interest as model systems for fundamental studies of size and interface effects in the nanosciences \cite{Erlebacher2001,Schofield2005,Weissmueller2009MRSBulletin}. They are also under consideration as materials basis for technologically relevant functional materials, for instance in actuation \cite{KramerNanoLetters2004,JinPtAu2010,Detsi2012,Zhang2016}, sensing \cite{ChenChen2011,Detsi2012,ZhangChen2013,Stenner2016,Seker2017}, energy storage \cite{Hou2014,Hou2016}, microfluidics \cite{Xue2014}, catalysis\cite{Zielasek2006,Ding2007,Zeis2008,DingChen2009,Fujita2012,Biener2015,FujitaUncercoordinated2016}, or lightweight structural materials \cite{Kato2011,KeWang2013,McCue2016,Okulov2017}. Almost invariably, such studies require precise data for the characteristic size of the struts or ligaments of the metal network. This characteristic structure size is an indispensable basis for discussing size effects and it implies key microstructural characteristics such as the volume-specific surface area or the abundance of low-coordinated atomic surface configurations. Size also controls mechanical strength \cite{Biener2005,Volkert2006,Hodge2007,YeJin2016,MamekaMRL2016} as well as fluid transport \cite{Xue2014}, and its link to the stiffness is under discussion \cite{MamekaMRL2016}. Actuation amplitudes \cite{JinPtAu2010} and signal strength or sensitivity in sensing \cite{Stenner2016} depend on the specific surface area, and the number of low coordinated sites is a key parameter for catalysis \cite{FujitaUncercoordinated2016}. Catalysis is also known to depend sensitively on the interatomic spacing at the surface and, thereby, on the lattice strain \cite{Mavrikakis2000,Kibler2005,Deng2014}. These remarks motivate a substantial interest in experimental techniques for quantifying the ligament size, the mean lattice parameter and the microstrain, that is, the variance of local lattice spacings.

X-ray powder diffraction is a classic approach to the above-mentioned task. The evolution of lattice parameters and of the coherent scattering length of nanoporous gold (NPG) during dealloying have been studied by that technique \cite{Schofield2008,Petegem2009,Toney2011}. Yet, so far there has been no dedicated analysis with respect to standard wide-angle X-ray patterns and its use as a microstructural characterisation method.

The classic approach to size effect on powder diffraction, as embodied in the Scherrer equation and in many later refinements, such as the Warren-Averbach and Williamson-Hall approaches and their variants \cite{KlugAlexander1974}, analyses scattering by a statistically relevant ensemble of very small crystalline particles in random orientation, for which the interparticle interference can be ignored. Dealloying-made nanoporous metals distinguish themselves from that scenario by one fundamental aspect: neighbouring ligaments are part of the same (porous) crystal and therefore scatter \emph{coherently}. Studies of millimetre-size nanoporous samples using ion challenging contrast \cite{Parida2006}, electron backscatter diffraction imaging \cite{Jin2009,YeJin2016}, X-ray Laue scattering \cite{Petegem2009}, and Bragg coherent X-ray diffractive imaging \cite{ChenWiegard2017} show that dealloying conserves the grain size of the initial polycrystalline aggregates which typically range between 10 to $\unit[100]{\rm \upmu m}$. The individual ligaments, typically 10 to \unit[100]{nm} in size, then share the same coherent crystal lattice. Even though microstructural defects that may degrade the lattice coherence -- specifically nonuniform strain fields \cite{SunGao2013,Ngo2015}, stacking faults \cite{Crowson2009,SunGao2013,Ngo2015}, and twins \cite{DouDerby2011,Ruestes2016} -- have been reported for nanoporous gold, the crystal size is clearly much larger than the ligament size.
This raises the question whether and how the apparent coherent scattering lengths -- as indicated by the Bragg reflection broadening -- relate to the ligament size of dealloying-made nanoporous metal.

Here, we present an X-ray powder diffraction study of dealloying-made nanoporous gold along with a discussion of the most elementary concepts of scattering by porous crystals. We performed virtual X-ray diffraction on atomistically modelled NPG in order to confirm that the contributions from nanoporosity and grain size are separately discernible if no relaxation of the structure was performed. Surprisingly, the contribution of the (large) grain size vanishes after thermal relaxation of the atomistically modelled structure. As in the simulation, the diffraction patterns of experimental NPG structures show no contributions from the grain size. We therefore employ samples that result from different dealloying and processing procedures to analyse their impacts on the diffraction patterns. The assessment of the results of our X-ray analysis is based on systematic scanning electron microscopy (SEM) data for ligament size and composition. We report and discuss the two following surprising findings: First, even though our theory suggests otherwise, experimental coherent scattering domain sizes as derived from Williamson-Hall analysis are strongly correlated to the "true" ligament sizes as derived by SEM. Second, the correlation between microstrain and Williamson-Hall ligament size in nanoporous gold is numerically for all practical purposes identical to what is found in a screening of other and differently derived nano\emph{crystalline} FCC metals. This latter finding is astounding since the microstrain should have quite different origin in nanoporous as compared to nanocrystalline metals.

\section{Theory: Scattering by Porous Crystals} \label{theory}

In wide-angle powder diffraction, the scattering vectors and the relative intensities of the Bragg reflections are determined by the \emph{atomic structure}, i.e. by the symmetry and the lattice parameter of the crystal lattice. The \emph{microstructure} can be taken, in the simplest approach, to affect the reflection breadth through size and microstrain effects. Unfortunately, the state-of-the-art on microstrain in nanoporous metal does not \emph{a priori} provide a basis for discussion, so that we will inspect that issue later, along with the experimental results of our work.
In contrast, the information on the geometry of dealloying-made metal network structures, as obtained by electron microscopy and small angle scattering, provide an excellent basis for discussing size effects, and this will be the sole focus of the present analysis. Our approach to the impact of porosity on the wide-angle powder diffraction pattern will be based on analogous treatments for isolated particles and nanocrystalline solids, see for instance the detailed description in \cite{Loeffler1995}.

\begin{figure*}[tb]
\begin{center}
\includegraphics[width=1.0\textwidth]{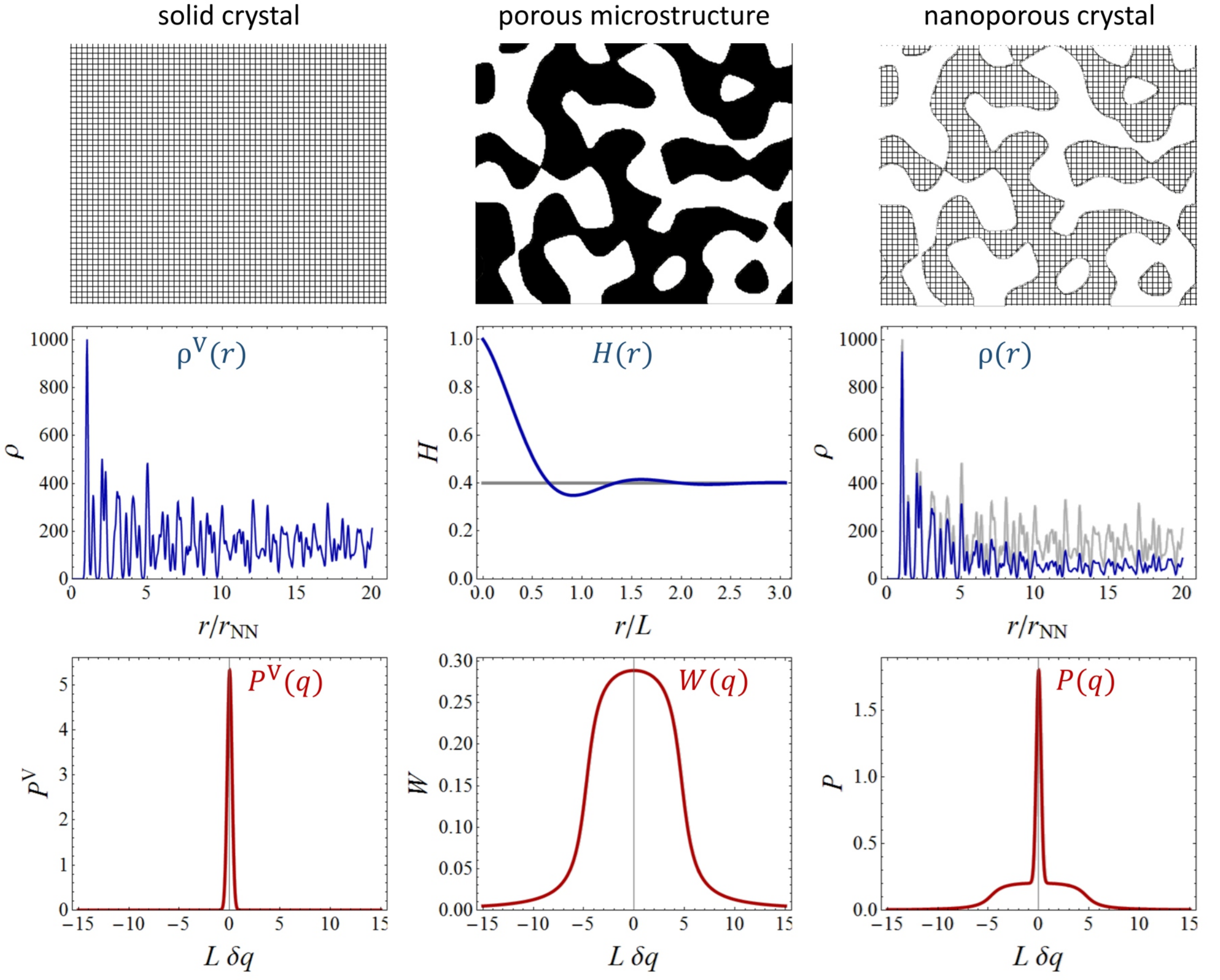}
\end{center}
\caption{Schematic representation of the various concepts involved in our discussion of the scattering of nanoporous crystals. Columns refer, from left to right, to the extended bulk crystal lattice, the geometry of the porous microstructure, and the nanoporous crystal. Rows refer, from top to bottom, to representation in real space, autocorrelated space and reciprocal space. For more details see the main text.}
\label{fig:Schema}
\end{figure*}

We analyse the diffraction by a polycrystalline aggregate  consisting of many randomly oriented crystallites (or ``grains''). Furthermore, each crystal contains pores, much smaller than the grain size. Because of the random orientation, (the structure is statistically isotropic and the interference function, $P$, depends on the scattering vector only through its magnitude, $q = 4 \uppi \sin \uptheta / \uplambda$, where $\uptheta$ denotes half the scattering angle and $\uplambda$ the wavelength.  Experimental diffraction data also depend on the atomic form factor, absorption, polarization, and incoherent scattering, all of which will be ignored for conciseness. We thus focus on $P$ which, for the isotropic problem, relates to the real-space structure through the radial distribution function (RDF) $\uprho(r)$ -- the orientation-averaged mean density of neighbouring atoms at distance $r$ from a mean central atom -- via
\begin{equation}
P(q)
=
4  \uppi  \int_0^\infty
 \left( \uprho(r) - \bar \uprho \right)
\frac{\sin q r }{q r }
\, r^2 \mathrm d r \, ,
\label{equation:P_interference}
\end{equation}
where $\bar \uprho$ denotes the mean atomic density, as derived e.g. in \cite{Sivia2011}. Figure~\ref{fig:Schema} shows schematic graphics of the functions which are discussed here.

The RDF of the massive (no pores) crystal lattice, $\uprho^{\sf V}(r)$, may be assumed known. This hypothetical massive lattice may contain defects such as dislocations, stacking faults and twin boundaries, which will be reflected in its RDF. A porous crystal may be constructed by cutting away atoms from the massive one. Since this process does not displace any of the remaining atoms, the peaks which represent the successive atomic neighbour shells in the RDF retain their positions but loose height. This loss is embodied by the microstructural envelope function $H(r)$ via
\begin{equation}
\uprho(r)
=
\uprho^{\sf V}(r) H(r)
\label{equation:rho_truncated}
\end{equation}
The envelope function is the autocorrelation function of a microstructural phase field $p(\mathbf x)$ which takes the values $p=1$ and $p=0$ for positions $\mathbf x$ in the solid phase and in the pore, respectively. It is useful to normalise $H$ to the volume, $V^{\sf S}$, of the \emph{solid phase} \cite{Loeffler1995}, so that
\begin{equation}
H(\mathbf r)
=
\frac 1 {V^{\sf S}}
\int p(\mathbf x) p(\mathbf x + \mathbf r )  \, \mathrm d^3 \mathbf x
\label{equation:envelope_definition}
\end{equation}
and specifically $H(0) = 1$.

As we are concerned with isotropic microstructures, we take $H$ to depend on the interatomic spacing $\mathbf r$ exclusively through its magnitude, $r$. The convolution theorem of Fourier transform implies that the interference function of the porous crystal is
\begin{equation}
P(q)
=
P^{\sf V}(q) \otimes W(q)
\label{equation:P_convolution}
\end{equation}
with $P^{\sf V}(q)$ the interference function of the massive crystal and with \cite{Loeffler1995}
\begin{equation}
W(q)
=
\frac 1 \uppi
\int_0^\infty H(r) \cos q r \,  \mathrm d r \,.
\label{equation:W_definition}
\end{equation}
Equations~\ref{equation:P_convolution} and \ref{equation:W_definition} imply that the effect of porosity is to broaden the very narrow diffraction peaks of the massive crystal by convolution with the cosine-transform of the envelope function. This, in turn, implies that discussing the scattering of porous crystals requires an analysis of the envelope function.

Autocorrelation functions such as $H(r)$ of equation~\ref{equation:envelope_definition} are best known from small-angle scattering studies, since the interference function near the origin of reciprocal space, $P^{\sf O}(q)$, is governed by the autocorrelation function of the scattering length density (see, e.g., Section 3.2 in reference~\cite{Michels2008}), which agrees with $H(r)$ except for a constant factor. Designating the scattering length density by $N$ and a characteristic scattering volume by $V^{\sf S}$, the small-angle scattering signal depends on $H$ as
\begin{equation}
P^{\sf O}(q)
=
4 \uppi
\frac {N^2} {V^{\sf S}}
\int_0^\infty
 H(r) \frac{\sin q r }{q r } \, r^2
\mathrm d r \, .
\label{equation:P_SAX}
\end{equation}

By definition, $H(0) =1$. For isolated particles, it is required that $H=0$ for distances larger than the longest intercept in the particle (the diameter, for the special case of a spherical particle). For the porous crystal, the value of $H$ will converge to that of the solid volume fraction, $\upvarphi$, at spacings much larger than the characteristic pore- or ligament size. The graph of $H(r)$ at intermediate distances $r$ depends on the degree of order in the pore or network structure. Dealloying-made network structures such as nanoporous gold exhibit significant order, as is evidenced by a pronounced peak in their small-angle scattering at $q$ in the order of the reciprocal ligament size \cite{Corcoran2003,Toney2011}.

As the microstructure of NPG resembles that of spinodally decomposed solutions \cite{Erlebacher2001,NewmanMRSBull1999,Farkas2007,SunGao2013,Ngo2015}, the Berk model for small-angle scattering by bicontinuous microstructures \cite{Berk1987} might provide an instructive approach \cite{Corcoran2003,Farkas2007}. The model works with the description of spinodal decomposition in terms of level sets in random superpositions of concentration waves \cite{CahnSpinodal1965} and achieves an excellent representation of the small-angle scattering interference peak in spinodal microstructures \cite{Berk1987,Berk1991}. The algebra of the Berk model is unwieldy and the expressions for $P(q)$ are poorly suited for the concise and qualitative discussion of the present case. Therefore, we introduce a much simpler expression as a toy envelope function that has no stringent relation to the geometry of spinodal or other bicontinuous microstructures while it reproduces important characteristics of their scattering.

Consider the following envelope function:
\begin{equation}
\tilde H(r)
=
\exp \left(- \frac r L \right)
\frac {2 L} {3 \pi r}
\sin \left( \frac {3 \pi r} {2 L} \right)
\, .
\label{equation:H_tilde}
\end{equation}
\begin{figure*}[tb]
\begin{center}
\includegraphics[width=1.0\textwidth]{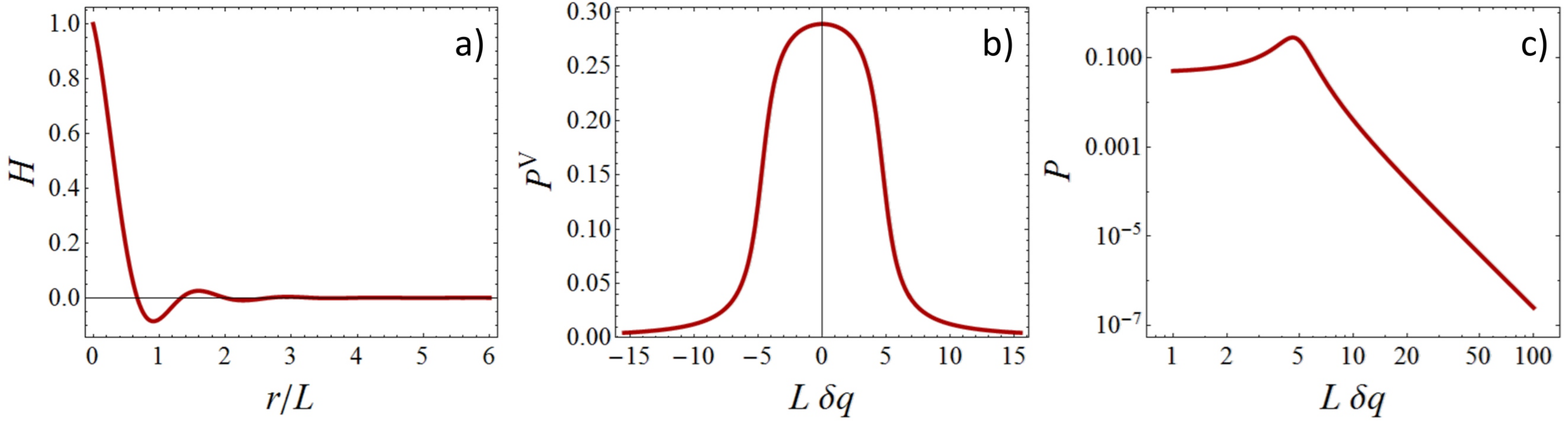}
\end{center}
\caption{Schematic representation of a) the toy envelope function $\tilde H(r)$ and of the corresponding b) wide-angle peak profile function $W(\delta q)$ and c) small-angle scattering intensity function $\tilde P ^{\sf O}(q)$.}
\label{fig:H_Tilde}
\end{figure*}
and its transforms
\begin{equation}
\tilde W(\delta q)
=
\frac L {3 \pi}
\left(
 \arctan  \left(  \frac{3 \pi} 2 - L \delta q \right)
 +
 \arctan  \left(  \frac{3 \pi} 2 + L \delta q \right)
 \right)
\, .
\label{equation:W_tilde}
\end{equation}
and
\begin{equation}
\tilde P^{\sf O}(q)
=
\frac {128 \pi}
{16 L^4 q^4+8 \left( 4 - 9 \pi ^2 \right) L^2 q^2 + \left( 4 + 9 \pi ^2 \right)^2}\, .
\label{equation:P_tilde}
\end{equation}
where we have taken $N = 1$ and, somewhat arbitrarily, $V^{\sf S}= L^3$.

We now inspect the properties of the three above toy functions depicted in figure~\ref{fig:H_Tilde}, bearing in mind that an approximation of the network structure is an array of cylinders with diameter (``ligament size'') $L$, characteristic spacing $d = 2 L$ between neighbouring ligaments and volume-specific surface area $\alpha$ given by
\begin{equation}
\alpha
=
4 / L \, .
\label{equation:alpha_cylindre}
\end{equation}
For use below we write the Scherrer formula \cite{Scherrer1918} in its form
\begin{equation}
\delta q_{\mathrm{size}}  = 2 \pi K / L
\label{equation:Scherrer}
\end{equation}    			
with $K$ the Scherrer constant and $L$ the relevant measure for size.

The following may be noted:
\begin{itemize}
  \item As would be expected for structures with a characteristic size $L$, the function $\tilde H(r)$ decays to near zero on the scale $r \rightarrow L$ (see figure \ref{fig:H_Tilde}a)).
  \item The initial slope of $\tilde H(r)$ is $L^{-1}$. This is consistent with the requirement that $\mathrm d \tilde H / \mathrm dr |_{r = 0} = -\alpha / 4$ for microstructures with discrete surfaces \cite{MelnichenkoBook2016} and with the volume-specific surface area of the idealised cylindrical ligaments, equation~\ref{equation:alpha_cylindre}.
  \item A minimum in $\tilde H(r)$ near $r = L$ (more precisely, at $r = 0.911 L$, see \ref{fig:H_Tilde}a)) agrees with the expectation for bicontinuous structures such as that in the central column of figure~\ref{fig:Schema}.
  \item The function $\tilde W(\delta q)$ has integral breadth of $1.73 \times 2 \pi / L$ and full width at half maximum of $1.53 \times 2 \pi / L$. Scherrer constants for equiaxed crystallites take values around 1, see for instance \cite{Markmann2008}. The present numerical values, 1.53 and 1.73, are larger but of similar order of magnitude. Within the scenario of our toy function, these values represent the Scherrer constants that relate the reflection breath to the ligament diameter.
  \item Asymptotically at large momentum transfer, $\tilde W(\delta q)$ varies as $(L/\delta q)^{-2}$, consistent with the Lorentzian behaviour expected by scattering from microstructures with discrete surfaces.
  \item The small-angle scattering function $\tilde P ^{\sf O}(q)$ has a pronounced interference peak at $q_{\rm I} = 1.47 \pi / L$.  The presence of the maximum agrees with observations for dealloying-made nanoporous gold and for spinodally decomposed microstructures, and its $q$-value agrees well with the estimate $q_{\rm I} = 1.23 \times 2 \pi / d$ for objects with a characteristic distance $d = 2L$.
  \item Asymptotically at large $q$, the small-angle scattering intensity varies as $\tilde P ^{\sf O}(q) = 2 \pi \alpha q^{-4}$. These are the expected power law and prefactor for a structure with discrete surfaces and specific surface area $\alpha$.
\end{itemize}

As an essential conclusion, the envelope function of the porous crystal can be separated into the sum of two contributions:
\begin{equation}
H(r)
=
(1- \upvarphi) \hat H(r) + \upvarphi
\end{equation}
where the information on the characteristic size and correlation of the ligament network is contained in the $r$-dependent part, $\hat H(r)$, whereas the constant represents the mean solid fraction. The former is expected to behave qualitatively as our toy function $\tilde H(r)$. As a consequence of equations~\ref{equation:W_definition} and \ref{equation:P_convolution}, the Bragg reflection shapes will then appear as schematically illustrated in figure \ref{fig:Schema}. The constant part of $H$, which transforms into a $\delta\text{-}$function, contributes a sharp central part of the reflection, whereas the $r$-dependent part contributes an additive peak, of width in the order of the reciprocal ligament size, which appears as a broad foot in each Bragg peak.

\section{Virtual Diffraction}

\subsection{Sample preparation and thermal relaxation}

As a verification of the theory in the previous Section, we performed a ``virtual''\cite{Markmann2008,Markmann2010} diffraction experiment, in which the powder diffraction pattern was computed based on an atomistic model of nanoporous gold obtained via molecular dynamics simulation. The diffraction pattern was then evaluated by the identical procedures as experimental data, see Section V below.

\begin{figure}[b]
	\begin{center}
		\includegraphics[width=0.8\columnwidth]{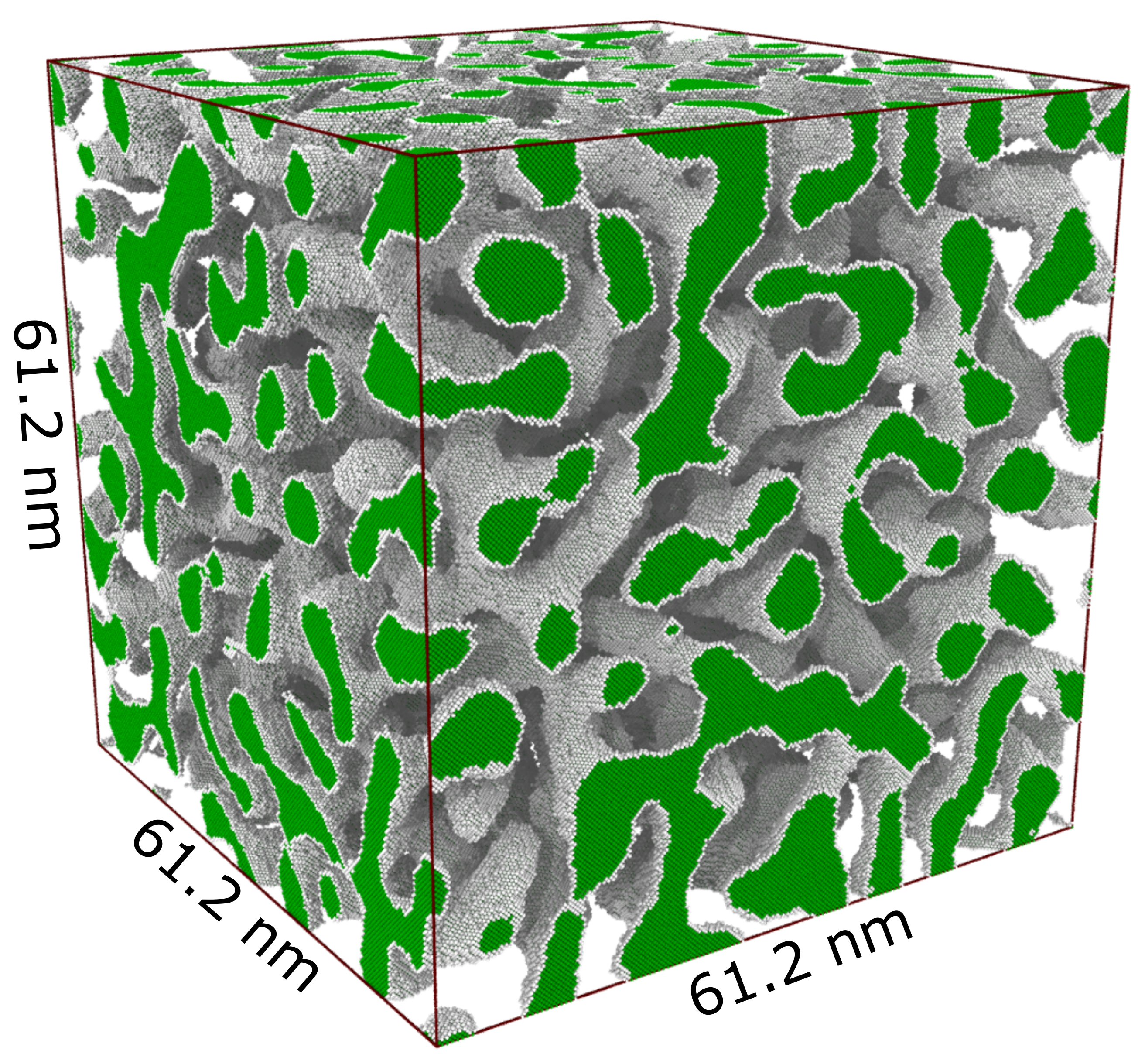}
	\end{center}
	\caption{Snapshot of the virtual nanoporous gold sample used in the simulation part of this study. The solid fraction of this structure is $0.2992$. For the sake of clarity, the colour code refers to surface atoms (white) and bulk atoms (green).}
	\label{fig:md_as_created}
\end{figure}

We created a nanoporous gold microstructure via imitating spinodal decomposition of a binary mixture on an FCC lattice, using the Metropolis Monte Carlo algorithm \cite{Newman1999}. Details of the simulation can be found in reference~\cite{Ngo2015}. Figure~\ref{fig:md_as_created} shows a snapshot of the as-created structure. The corresponding simulation box has the size of $150 \times 150 \times 150$ lattice spacings ($61.2 \times 61.2 \times 61.2$ nm) and $\braket{100}$ edges. The solid fraction of this structure is $0.2992$. The alpha-shape surface reconstruction algorithm (with a probe radius of $3$~\AA ) \cite{Stukowski2014} gives the ratio of surface area per solid volume $\alpha=0.997/\mathrm{nm}$. According to equation~\ref{equation:alpha_cylindre}, assuming cylindrical ligaments, this corresponds to a ligament size of \unit[4]{nm}.

Molecular dynamics (MD) simulations were carried out with the simulation code \textsc{Lammps} \cite{Plimpton1995}, using an embedded-atom method potential for gold \cite{Foiles1986}. First, the atom positions were athermally relaxed via an energy minimization using the conjugate gradient algorithm. The relative change in energy and force tolerance at convergence were less than $10^{-12}$ and $10^{-4}$ eV/\AA, respectively. Then, the virtual sample was thermally relaxed for 1~ns at 300~K under zero external load. Periodic boundary conditions were applied in all simulations.

Two states of the model were investigated in the virtual diffraction experiment: Firstly, the nanoporous structure in its unrelaxed state, that is, the truncated ideal lattice. This represents an idealised porous crystal as it is created by the Monte Carlo algorithm. All atoms occupy sites of the periodic crystal lattice and there are no strain and no thermal displacement. This structure emulates the one investigated in the theory of Section II. Secondly, the nanoporous structure in its relaxed state. The structure is represented by a snapshot of the configuration that emerges from the relaxation by MD as described above. It incorporates static lattice strain as well as the instantaneous atomic displacement due to thermal vibrations at the instant of the snapshot.

\begin{figure}[b]
	\begin{center}
		\includegraphics[width=1.0\columnwidth]{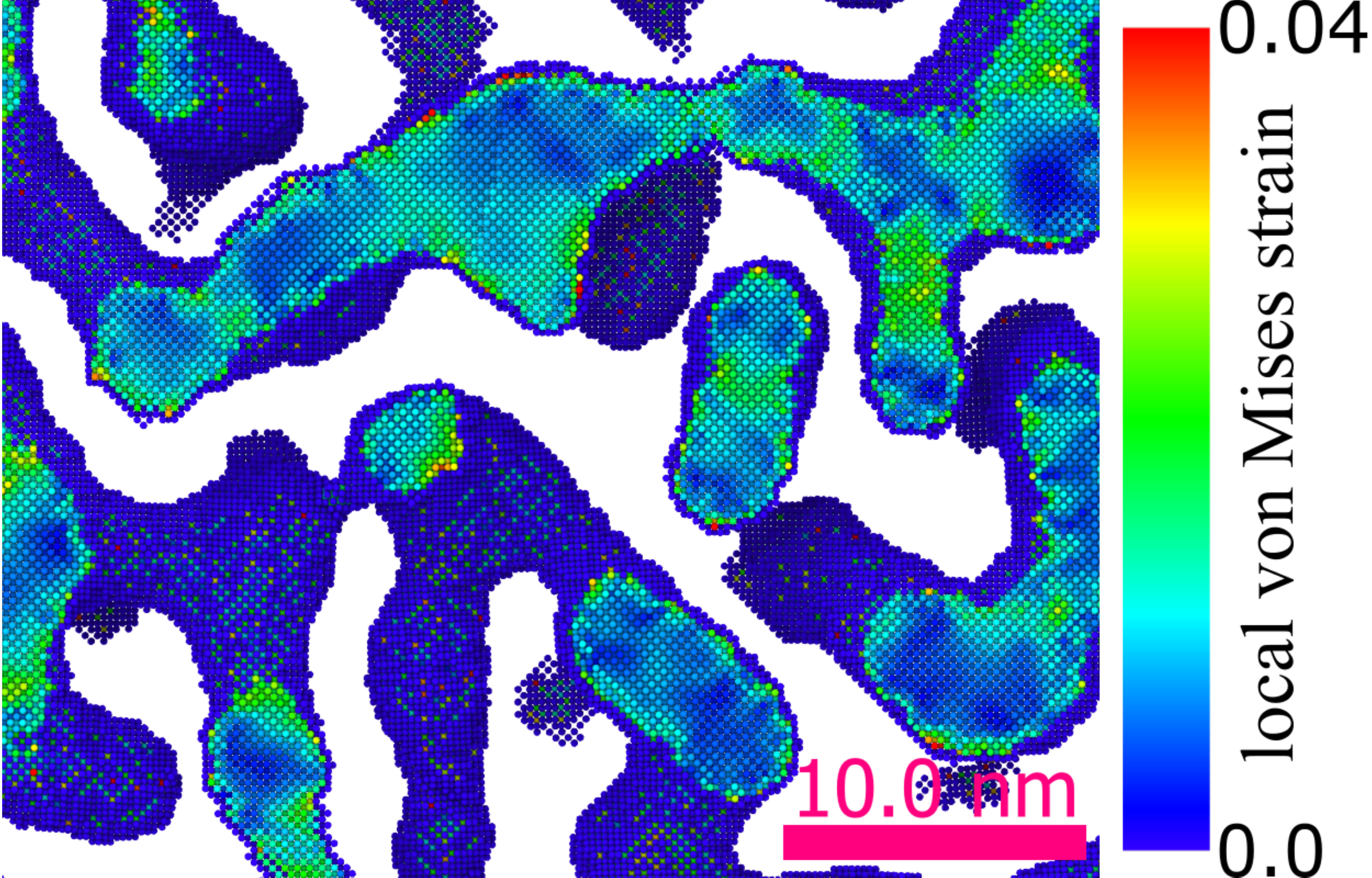}
	\end{center}
	\caption{Von Mises strain map of a cross-section in $\langle 100 \rangle$ direction inside the simulation cell after relaxation and quenching (for details see text). The elastic tensor components were determined according to \cite{Stukowski2012}. The von Mises strain was computed with the formula given in \cite{Shimizu2007}.}
	\label{fig:md_strainmap}
\end{figure}

Figure \ref{fig:md_strainmap} shows the amount of local von Mises strain in a colour coded map cut along a crystallographic $\langle 100 \rangle$ plane. In order to eliminate contributions of the thermal vibrations to the local strain of the atoms, the structure was, after its relaxation at \unit[300]{K}, quenched at \unit[0.01]{K} followed by an additional relaxation for \unit[1]{ns} at \unit[0.01]{K}.
Strain gradients inside the ligaments are clearly discernible even far from the surface. Regions of high and low strain divide the volume into regions of a size which correlates to the ligament size.

\subsection{Diffraction pattern simulation}

The calculation of the X-ray diffraction used the procedures presented in reference~\cite{Markmann2008} using wavelengths $\lambda_1=\unit[1.54056]{\AA}$ and $\lambda_2=\unit[1.54439]{\AA}$ with an intensity
ratio of 2:1 for Cu K$_{\rm \alpha}$ radiation and the atomic form factor of gold \cite{Wilson1995}.

This kind of virtual diffraction on atom positions as calculated by molecular dynamics simulation or density functional theory is an established procedure and was done in similar ways in \cite{Derlet2005,Spearot2014}.
Basically, we calculate a powder pattern of cuboid particles, which are identical to the MD simulated nanoporous gold. This represents well the situation in the lab experiment described in the following section where coarse grained polycrystalline gold is penetrated by a nanoporous network. The simulation box represents a single grain and the use of the Debye formula to calculate the diffraction pattern (compare equation~(\ref{equation:P_interference}) and ref~\cite{Markmann2008}) perfectly mimics the polycrystalline character of the samples.
\begin{figure}[b]
	\begin{center}
		\includegraphics[width=1.0\columnwidth]{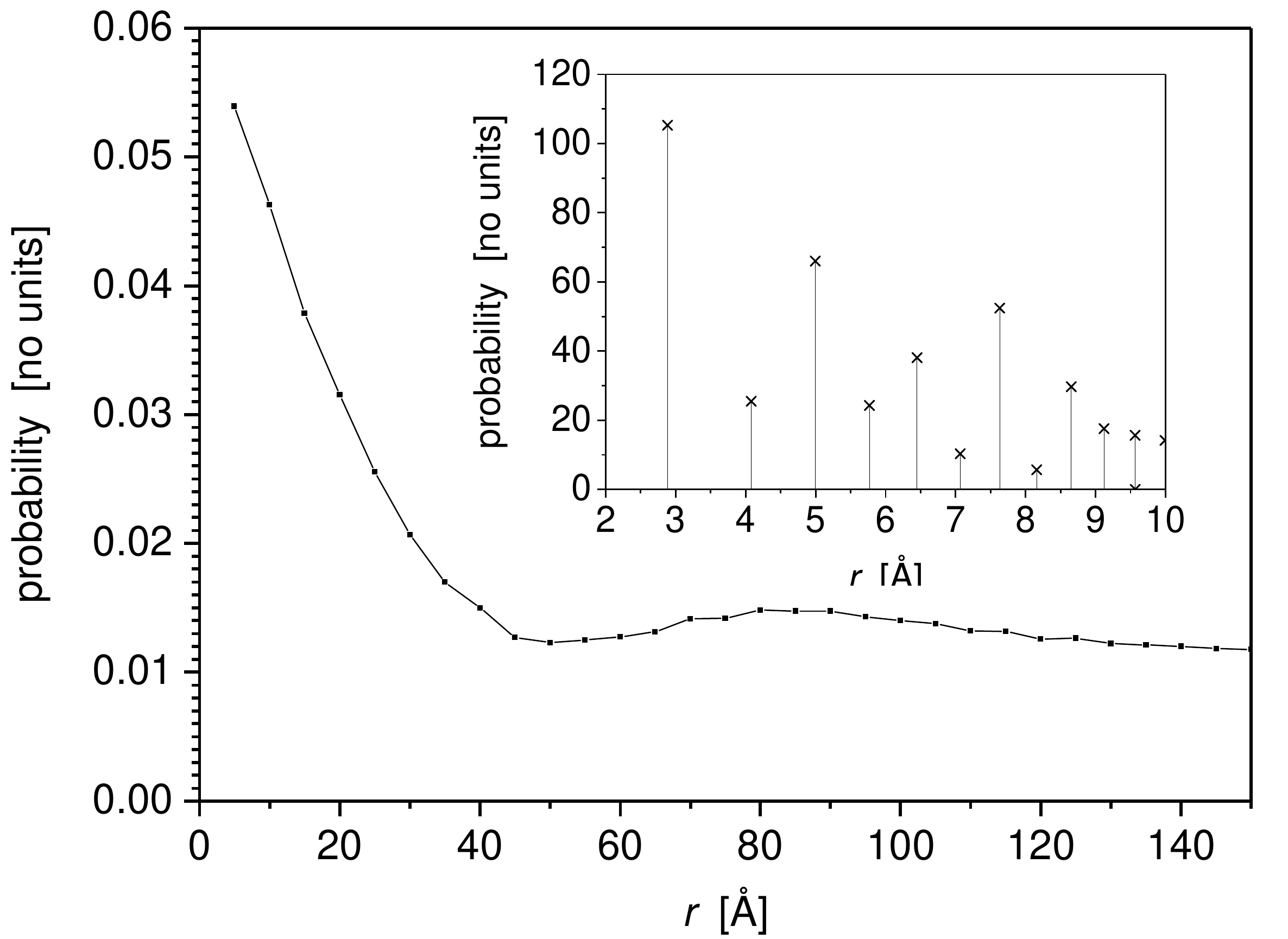}
	\end{center}
	\caption{Radial distribution function of a nanoporous structure where gold atoms are placed on perfect FCC positions summed up in a histogram with a sampling size of \unit[5]{\AA}. The inset shows the same function for small distances without the averaging.}
	\label{fig:rdf_ascreated}
\end{figure}
The first step is the determination of the orientation-averaged radial distribution function (RDF). This function is discontinuous for the as-constructed sample, with non-zero values only in the discrete neighbouring atom shells. This is shown in the inset of figure~\ref{fig:rdf_ascreated}. The reorganization of the RDF into a coarse histogram with a grid size of \unit[5]{\AA} results in a smooth function looking strikingly similar to the one shown in the centre of figure~\ref{fig:Schema} of the nanoporous microstructure without the crystalline contribution.

In contrast to the situation described in the preceding paragraph, we do not use periodic boundary conditions when we calculate the RDF. This means the RDF does not reach a constant value corresponding to the mean solid fraction but decays to zero for values larger than the box diagonal. This second envelope function is necessary because it provides a smooth transition for the Fourier transform necessary to calculate the scattering function. Otherwise, heavy oscillations are produced by the calculation. But size of the MD box is one order of magnitude larger than the ligament size. Therefore, this additional envelope function does not influence the broad part of the peaks but only the sharp peaks representing the scattering signal of the crystalline host material. For the thermally relaxed structure, this additional broadening can be readily neglected.

\begin{figure}[tb]
	\begin{center}
		\includegraphics[width=1.0\columnwidth]{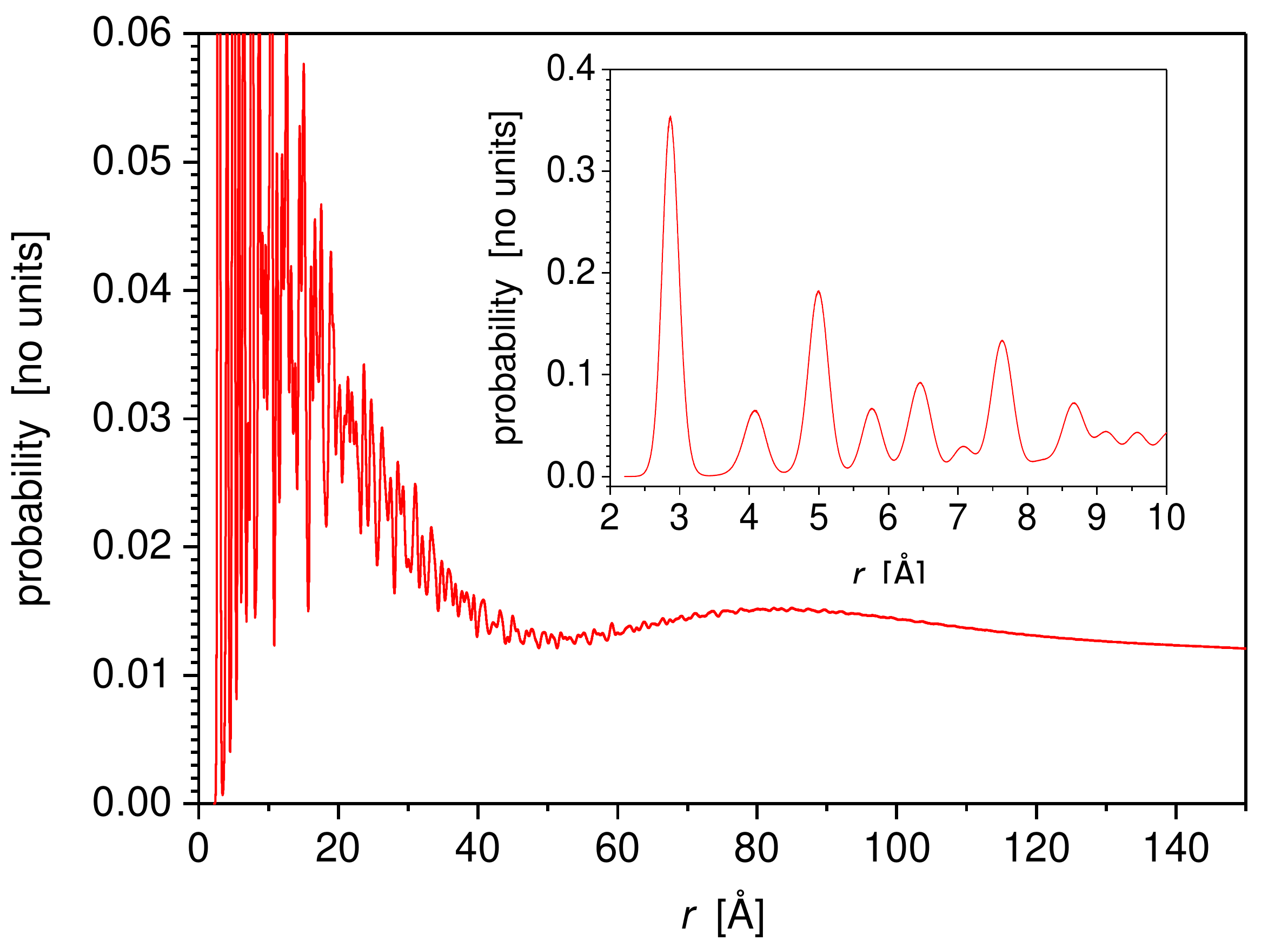}
	\end{center}
	\caption{Radial distribution function (RDF) of a nanoporous structure after relaxation at \unit[300]{K} for \unit[1]{ns} by molecular dynamics simulation. Thermal vibrations turn the calculated RDF into a continuous function showing the occupation of the neighbouring shells (inset). The RDF in the range of up to \unit[150]{\AA} is shown on the original fine grid.}
	\label{fig:rdf_annealed}
\end{figure}

The RDF for a snapshot of the relaxed sample is shown in figure~\ref{fig:rdf_annealed}. As a consequence of the relaxation,, the discrete shells have been replaced by broadened peaks.

\begin{figure}[tb]
	\begin{center}
	\includegraphics[width=1.0\columnwidth]{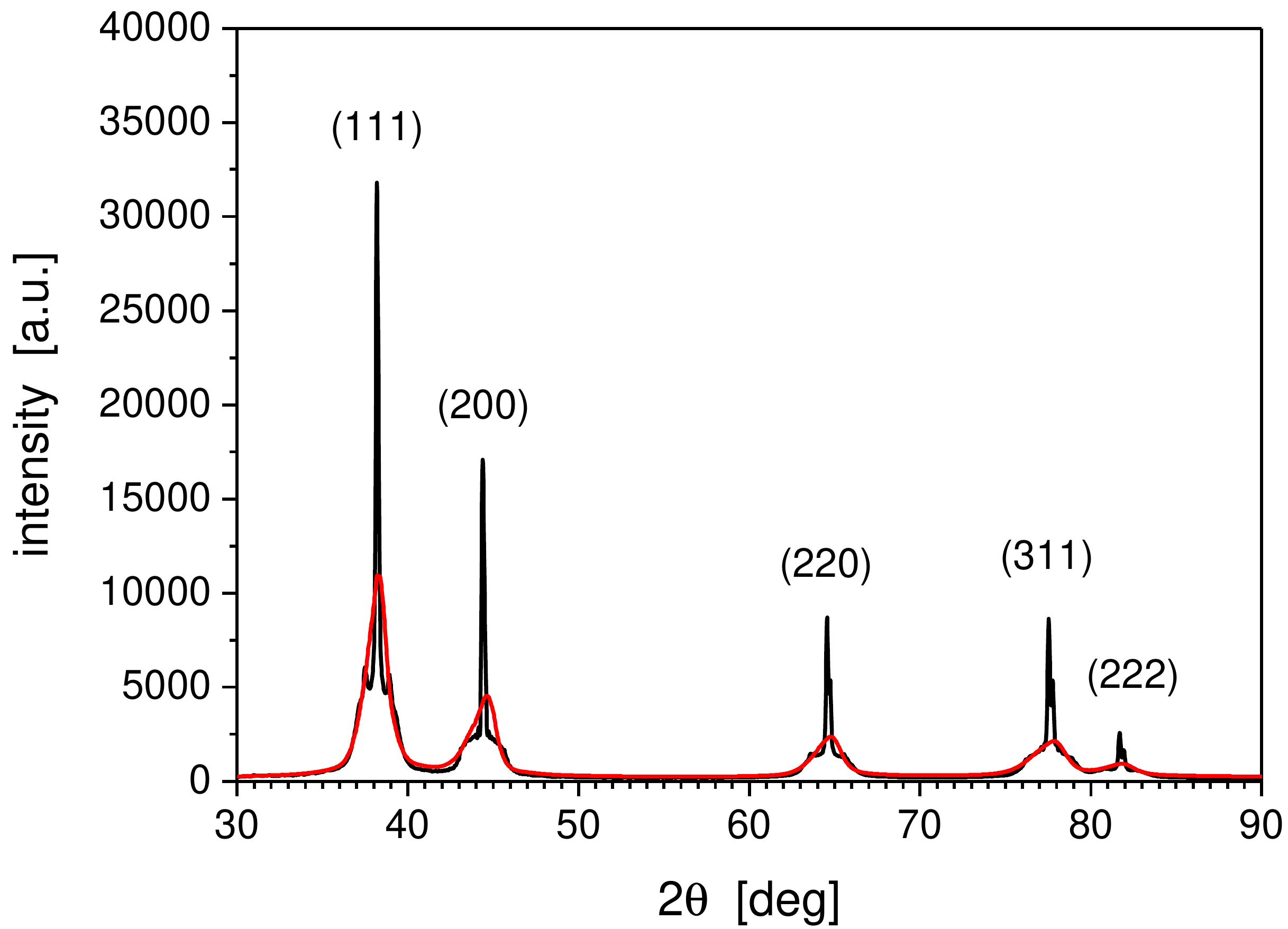}
	\end{center}
	\caption{Virtual diffraction patterns of the nanoporous structure in the unrelaxed state (black) and after thermal relaxation at \unit[300]{K} for \unit[1]{ns} by molecular dynamics simulation (red).}
	\label{fig:diffpatterns}
\end{figure}

The next step was the calculation of scattered intensity versus diffraction angle from the RDFs according to the procedure in \cite{Markmann2008}. Figure~\ref{fig:diffpatterns} compares the diffraction patterns for the unrelaxed and relaxed samples.
In view of the considerations in Section II it is remarkable that the unrelaxed sample indeed shows Bragg reflection superimposing a narrow central component and a broad basis. This is particularly evident for the (200) and (311) peaks. The (111) and (220) peaks exhibit satellite maxima, displaced by $\sim 0.6 - 0.7^\circ$ on the $2\theta$ scale from the peak centre. The angle shift corresponds to a distance of roughly \unit[12 - 14]{nm} in real space, much larger than the ligament size of \unit[4]{nm}. This means, these satellites most probably are higher order reflections of the simulation box itself.

The red line in figure~\ref{fig:diffpatterns} represents the virtual diffraction pattern of the relaxed sample. It is seen that the sharp central peak has been lost and that uniformly broadened reflections have emerged. We emphasize that the broadening cannot be the signature of thermal disorder, since that signature would not be broadening but a reduction in the Debye-Waller factor or, in other words, a reduction in peak height \cite{Krivoglaz1969}.
Therefore, the broadening can be attributed to a loss of long-range lattice coherency. This effect may be attributed to either a reduction in the coherently scattering domain size or to microstrain. For the relaxed sample, both values can be determined from the dependency of the peak width on the peak position. The analysis procedure is described in Section IV below. Here, it suggests a domain size of $\unit[7.8\pm0.4]{nm}$ and a microstrain of $0.61\pm0.07\%$.

\section{Experimental Procedures}

\subsection{Preparation of nanoporous samples \label{synthesisroutes}}

Ag-Au alloys, at ratio, Ag$_{75}$Au$_{25}$ (subscripts in atomic $\%$), were produced by arc-melting the pure metals. A thermal annealing with the ingot sealed in fused silica under Ar was performed for $\unit[5]{d}$ at $\unit[925]{^{\circ} C}$ in order to homogenise the Ag-Au alloy by bulk diffusion.
The following step-wise cold-rolling procedure (each step to $\unit[60]{\%}$ reduction of the initial thickness) of the alloy ingot (interrupted by a $\unit[5]{min}$ thermal relaxation period at $\unit[650]{^{\circ} C}$) achieved a foil thickness of $\unit[150]{\rm \mu m}$. For samples denoted as ``Alloy, cold-worked'', the recovery after the last rolling step was omitted. Afterwards, all foils were laser-cut to give circular discs of $\unit[5]{mm}$ diameter.

Selective Ag dissolution (chemically and electrochemically) created NPG samples. We used three different dealloying protocols as exposed in details in reference~\cite{Graf2017}. In brief:

\begin{itemize}
  \item Route \textbf{A}: dealloying in $\unit[1]{M}$ $\ce{HClO4}$ against a coiled $\ce{Ag}$ wire as counter electrode (CE) under a stepped dealloying potential regime (potentials were $\unit[1050]{mV}$ for $\unit[24]{h}$, $\unit[1100]{mV}$ for $\unit[8]{h}$, $\unit[1150]{mV}$ for $\unit[8]{h}$ and $\unit[1200]{mV}$ for $\unit[10]{h}$ vs. a $\ce{Ag} / \ce{AgCl}$ reference electrode (RE)).

  \item ``\textbf{Fast dealloying}'' refers to samples made by Route A but at constant and unusually positive dealloying potential in order to achieve a brute force, very fast corrosion. Dealloying potentials were $\unit[1200]{mV}$, ``$\unit[1400]{mV}$'' or ``$\unit[1600]{mV}$''. The dealloying was here more strongly in the oxygen species adsorption regime, so that surfaces of the samples are likely covered by adsorbed oxygen species or oxide.

  \item Route \textbf{B} involved potentiostatic dealloying according to Wittstock et al. \cite{Wittstock2011} in $\unit[5]{M}$ $\ce{HNO3}$ against a $\ce{Pt}$ CE under a constant dealloying potential of $\unit[60]{mV}$ for $\unit[24]{h}$ vs.~a $\ce{Pt}$ pseudo-RE.

  \item Route \textbf{C} samples were produced under open-circuit conditions according to Rouya at al. \cite{Rouya2012}, i.e. by immersion into thermostated ($\unit[65]{^{\circ}C}$) semi-concentrated ${\ce{HNO3}}$ for $\unit[24]{h}$.
\end{itemize}

Further NPG modifications were achieved using electrochemical and thermal treatments, i.e. samples AD(A) (as produced by route A) that are noted with ``$\unit[-500...+800]{mV}$'' or ``$\unit[-500...+1300]{mV}$'' underwent an additional electrochemical treatment, i.e. potential cycling (CV) within the respective potential window for $\unit[15]{cycles}$ at $\unit[5]{mV/s}$. Samples AD-A noted with ``$\unit[10]{min}$'' or ``$\unit[6]{h}$'' underwent an additional thermal treatment at $\unit[300]{^{\circ}C}$ for the respective duration under air.

All electrochemical processes were controlled by a potentiostat (Metrohm Autolab PGStat10).

Scanning electron microscopy (SEM) was performed to estimate the bulk ligament size, $L_{\rm SEM}$, after sample fracture and cross-view recording at a Leo Gemini 1530 equipped with an energy-dispersive X-ray spectroscopy (EDS) sensor by Oxford instruments.
$L_{\rm SEM}$ was estimated by averaging 20 diameters readings and taking the standard deviation as the absolute error.

\subsection{Diffraction}

X-ray diffraction used a powder diffractometer (Bruker D8 Advance) in $\uptheta$-$\uptheta$ focusing Bragg-Brentano geometry, with Ni-filtered Cu $\rm K_\upalpha$ radiation and a linear position-sensitive detector (LynxEye). Bragg reflection parameters were obtained from fits by $\rm K_{\upalpha_{1/2}}$ doublets of split Pearson VII functions.

Average lattice parameters were refined using the appropriate variant of the Nelson-Riley approach along with a correction for stacking fault displacements. The algorithm also supplied an estimate of the stacking-fault probability, see details in reference~\cite{Markmann2003,Markmann2010}. The Williamson-Hall analysis assumed Gaussian strain and Cauchy-type size broadening \cite{KlugAlexander1974}, following the procedures of reference~\cite{Markmann2008} which were validated in virtual experiments based on MD-generated samples of nanocrystalline metal \cite{Markmann2008,Stukowski2009}. The instrumental line broadening, as obtained using the NIST Standard Reference Material SRM 660b, was corrected for.

\section{Results and Analysis}

\subsection{Sample characterization}

\begin{figure}[tb]
\begin{center}
\includegraphics[width=1.0\columnwidth]{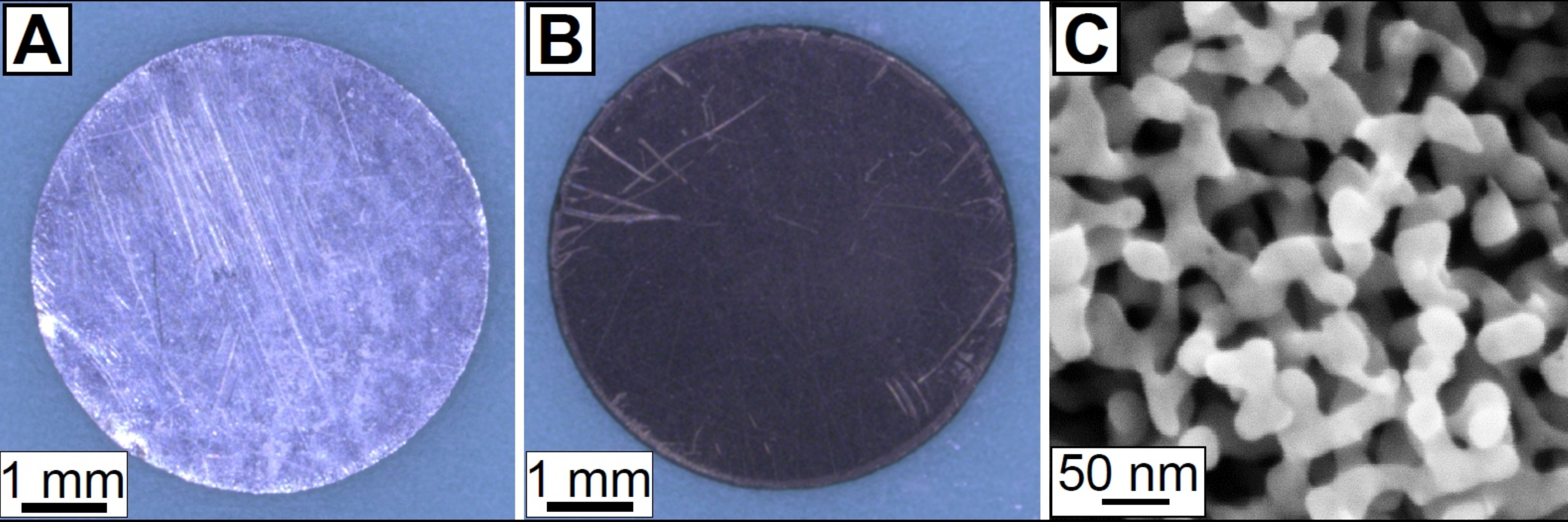}
\end{center}
\caption{Alloy and nanoporous samples to be characterised by X-ray diffraction: Top-view photographs of Ag$_{75}$Au$_{25}$ alloy discs before (\textbf{A}) and after dealloying (\textbf{B}). \textbf{C}: SEM image of a nanoporous gold sample as dealloyed by route A (see text for details) and prepared by sample fracture, yielding a ligament size, $L_{\rm B}=\unit[(21 \pm 4)]{nm}$.}
\label{fig:Samples}
\end{figure}

Figure~\ref{fig:Samples} illustrates the sample geometry and the microstructure. Part A shows the master alloy while part B shows a nanoporous sample after dealloying by Route A. The nanoporosity is seen in the scanning electron micrograph, part C. The ligament size, $L_{\rm SEM}$, for this example was determined as $\unit[21 \pm 4]{nm}$. Nanoporous gold samples made by dealloying contain residual silver. The residual silver atom fraction, $x_{\rm Ag}$, monotonously decreases with increasing ligament size [53]. Table ~\ref{silvercontents} indicates $x_{\rm Ag}$ values for the samples of the present study.

\begin{table}[tb]
	\caption{Silver fraction of the master alloy and the nanoporous gold samples. For sample denomination see Section~\ref{synthesisroutes}.
		 \label{silvercontents}}
	\begin{tabularx}{\columnwidth}{X|X}	
		Material &  Silver fraction [at\%] \\
		\hline \\[-0.9ex]
		Master alloy & \hspace{3em} 72.0 \\
		AD(A) &  \hspace{3em} 6.4\\
		AD(A) -500...\unit[+800]{mV} &  \hspace{3em} 1.3 \\
		AD(A) -500...\unit[+1300]{mV} &  \hspace{3em} 2.2\\
		AD(A) \unit[10]{min} &  \hspace{3em} 3.8 \\
		AD(A) \unit[6]{h} &  \hspace{3em} 3.4 \\
		AD(B) &  \hspace{3em} 1.8 \\
		AD(C) &  \hspace{3em} 8.6 \\
		AD(A) \unit[1200]{mV} &  \hspace{3em} 14.6 \\
		AD(A) \unit[1400]{mV} &  \hspace{3em} 5.2 \\
		AD(A) \unit[1600]{mV} &  \hspace{3em} 8.2 \\
	\end{tabularx}
\end{table}

\subsection{Experimental diffraction patterns and data analysis}

\begin{figure}[tb]
\begin{center}
\includegraphics[width=1.0\columnwidth]{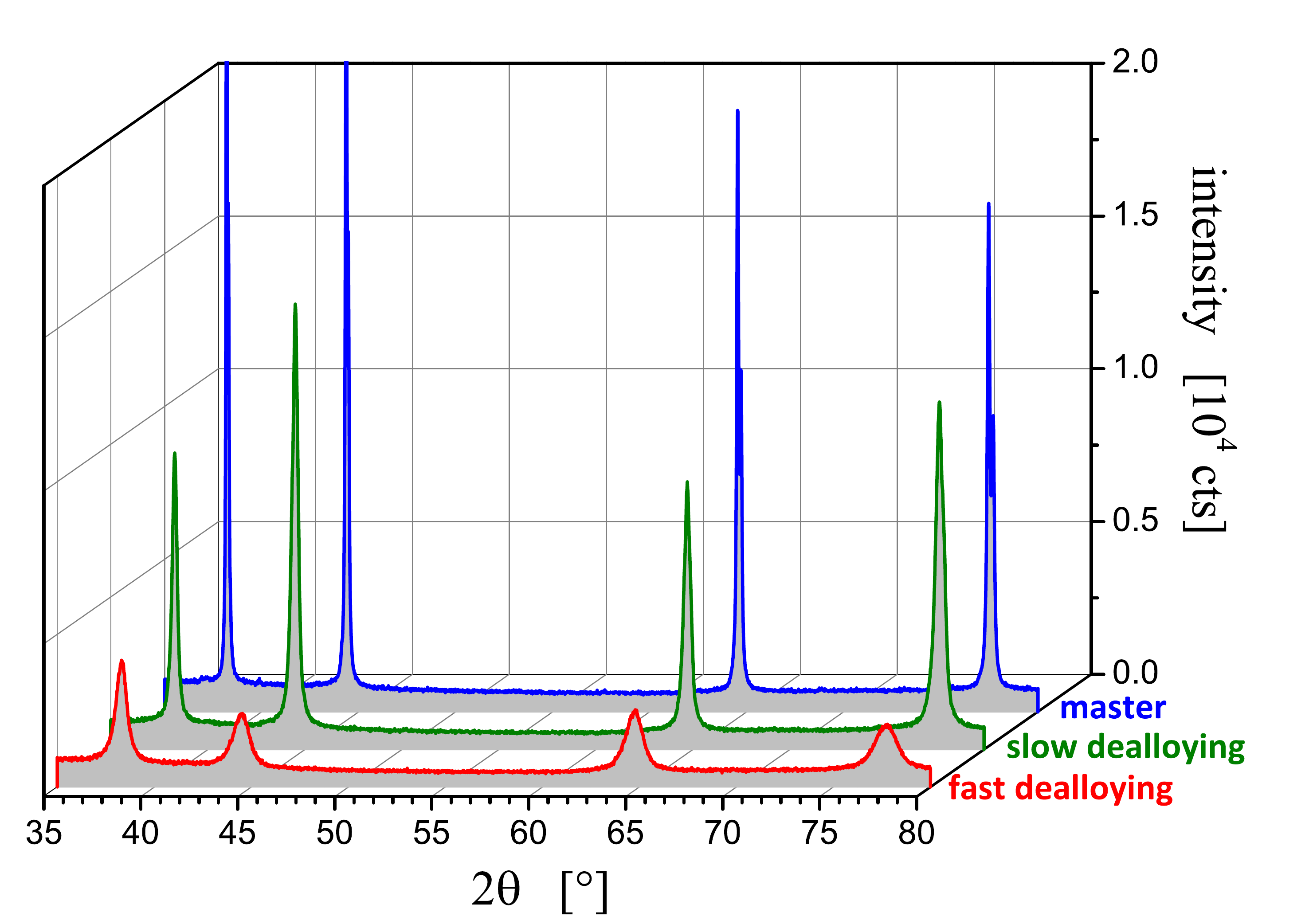}
\end{center}
\caption{Diffraction patterns of scattering intensity versus diffraction angle $2 \theta$ for the thermally recovered $\ce{Ag75Au25}$ master alloy and for two nanoporous gold samples prepared using different dealloying protocols (see main text for details). In order to convey reflection profiles, only part of the experimental angle range ($30^{\circ}$ to $130^{\circ}$) is shown.}
\label{fig:diffpattern}
\end{figure}

Diffraction patterns were recorded in the angle range $30^{\circ}$ to $130^{\circ}$, covering the Bragg reflections up to (420).
Figure~\ref{fig:diffpattern} exemplifies the results. All peaks correspond to the FCC crystal lattice of gold. As compared to the master alloy (blue), the peaks of NPG made by slow dealloying (green) are broadened, and fast dealloying (red) leads to even stronger broadening.  Similar results were obtained for all samples.

The relative intensities of (111) and (200) reflections varied between samples; we attribute this to statistical variation arising from the small number of irradiated crystallites.
As the most notable observation in the context of our discussion, the reflection profiles of the dealloyed samples appear uniformly broadened; in no case did we observe the bimodal peak profile which is predicted for the strain-free porous crystal with a truncated lattice (as in the right column of figure~\ref{fig:Schema}).
\begin{figure}[tb]
\begin{center}
\includegraphics[width=0.8\columnwidth]{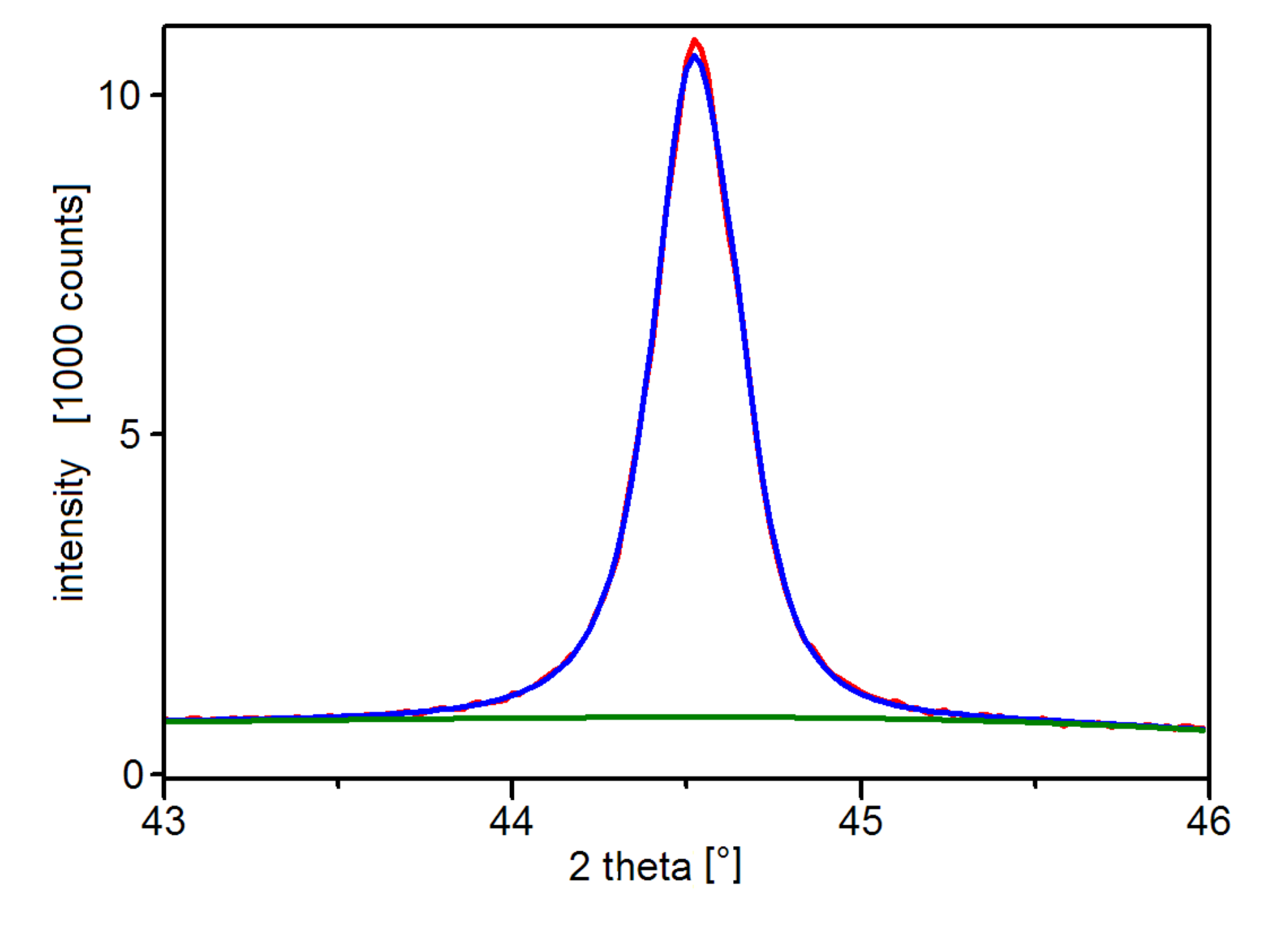}
\end{center}
\caption{Example for peak profile fitting, here for the (200) reflection,  here for a sample dealloyed in $\rm \ce{HClO4}$ for 50~h and with 38 nm ligament size (sample AD(A)-500...$\unit[+800]{mV}$). The fit with $\rm K_{\upalpha_{1/2}}$ doublets of split Pearson VII functions function (blue) in comparison with the experiment (red; graph mostly covered by the fit line). Green line is a background fit.}
\label{fig:profile_fit}
\end{figure}

As a basis for the evaluation of lattice parameters and for the Williamson-Hall analysis, all Bragg reflections in each data set where fitted with $\rm K_{\upalpha_{1/2}}$ doublets of split Pearson VII functions. Figure~\ref{fig:profile_fit} exemplifies the typical quality of the fit, here for the (200) reflection of a slowly dealloyed sample with $L_{\rm SEM}=\unit[38]{nm}$. The fit function is seen to provide a good representation of the experimental data, except for a small deviation near the peak maximum. Similar results could be obtained in each case.

\begin{figure}[tb]
\begin{center}
\includegraphics[width=0.8\columnwidth]{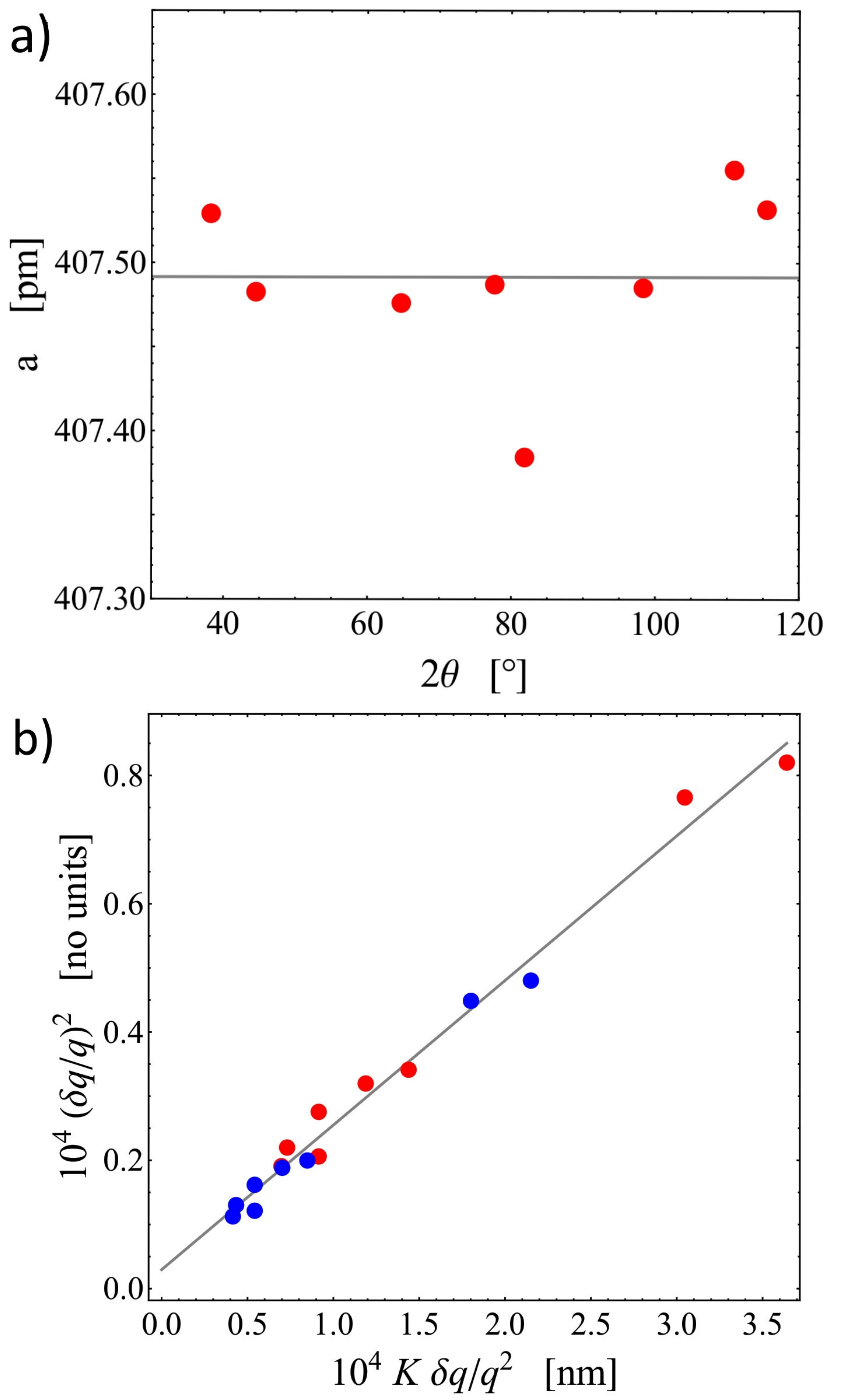}
\end{center}
\caption{Example for the analysis of diffraction data, here for a sample dealloyed in $\rm HClO_4$ for 50~h (sample AD(A)).
\textbf{a)} Single-peak lattice parameters (red circles) after Nelson-Riley correction for all Bragg reflections up to (420). Refined lattice parameter (grey line) is $a = 407.49 \pm 0.05 \rm pm$. Note that the deviations of single-peak lattice parameters are also used to determine stacking fault densities.
\textbf{b)} Modified Williamson-Hall plot of $K(\delta q/q)^2$ versus $\delta q/q^2$ with $q$ the scattering vector magnitude, $\delta q$ the reflection width and $K$ the appropriate Scherrer constant for integral breadth (red circles, $K = 1.075$) and full width at half maximum (blue circles, $K = 0.830$). Linear regression jointly to both data sets implies an apparent grain size $D = \unit[37.2\pm1.2]{nm}$ and a root-mean square microstrain $\langle \varepsilon^2 \rangle^{1/2} = (0.086 \pm 0.016) \%$.}
\label{fig:WH}
\end{figure}

The evaluation of lattice parameters was based on the maximum intensity positions from the fit. Figure~\ref{fig:WH}a) exemplifies the single-peak lattice parameters after correction for height misalignment and stacking-fault induced peak shifts \cite{Markmann2003}; the refined lattice parameter and its confidence intervals were determined as the mean and variance of such data sets.
Figure~\ref{fig:WH}b) exemplifies the modified (Cauchy/Gauss) Williamson-Hall analysis; combinations of peak width and position in $q$-space are plotted so that the slope of the straight line of best fit scales with the apparent grain size while the ordinate intercept measures the microstrain. We analysed both, the integral reflection breadth (red data points) and the full width at half maximum (FWHM, blue data points), applying separate Scherrer constants $K$ as established in reference~\cite{Markmann2008}. It is seen that the two separate measures for the peak broadening provide consistent results, and that the data agrees well with the straight-line behaviour expected for a combination of Cauchy-type size broadening and Gaussian strain broadening. The apparent grain size, $D$, and microstrain, $\langle \varepsilon^2 \rangle^{1/2}$, are obtained from the straight line of best fit to the data. In the example, they are $D = \unit[37.2\pm1.2]{nm}$ and $\langle \varepsilon^2 \rangle^{1/2} = (0.086 \pm 0.016)\%$.

\subsection{Lattice parameter}

\begin{figure}[tb]
\begin{center}
\includegraphics[width=0.8\columnwidth]{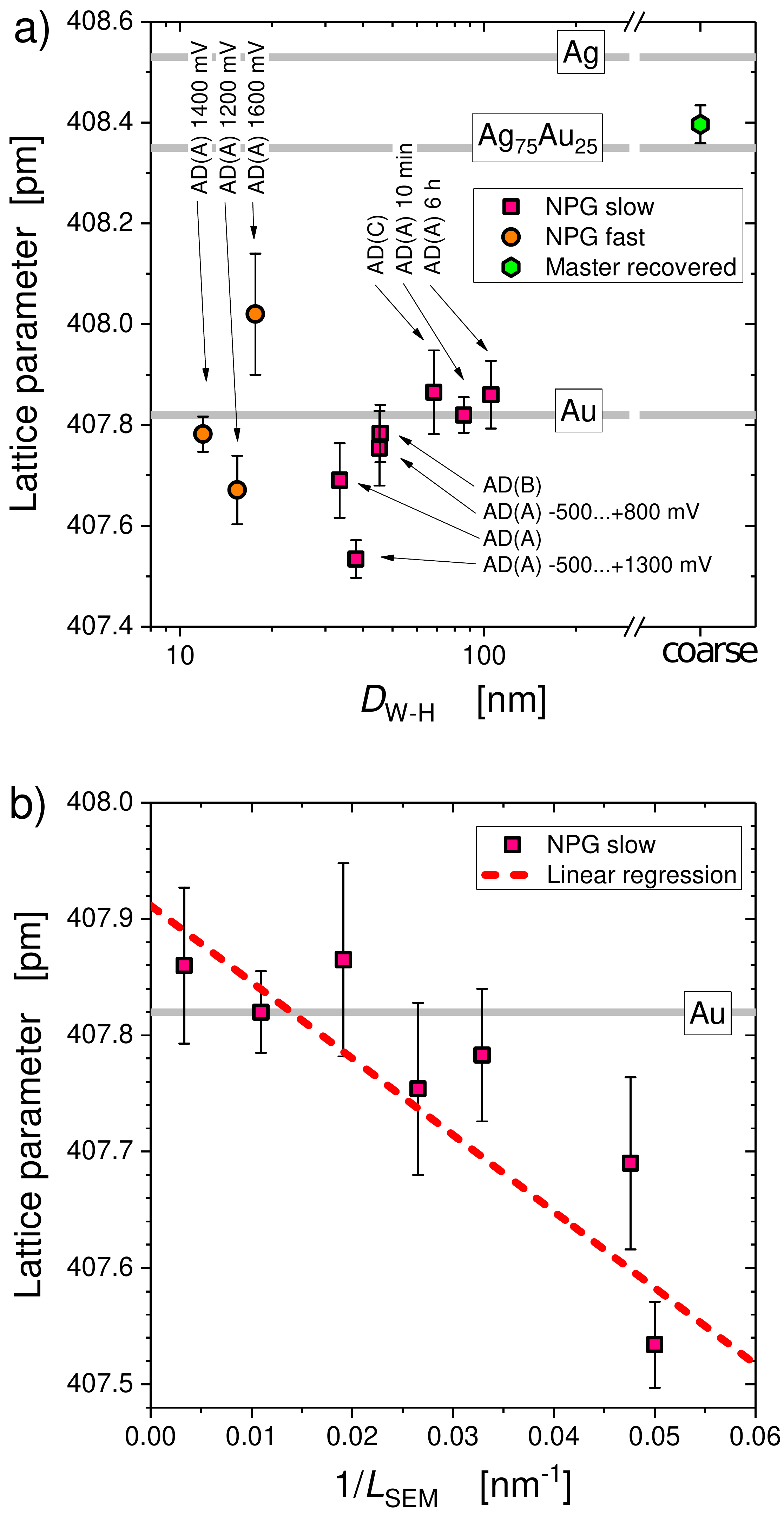}
\end{center}
\caption{ \textbf{a)} Lattice parameters, $a$, of the thermally recovered master alloy before and after dealloying by various protocols and subsequent post-treatments (see text for details) in dependence of the Williamson-Hall parameter $D_{\rm W-H}$. Horizontal grey lines represent the theoretical $a$ value of pure Au, Ag and the master alloy, respectively. Error bars correspond to the evaluated standard errors from the straight line fits (Nelson-Riley). See text for sample assignment.  \textbf{b)} shows the lattice parameter plotted vs.\ the inverse of the bulk ligament size, $L_{\rm SEM}$, as derived from SEM images, for the set of slowly dealloyed NPG as well as a linear regression of that part of the data.}
\label{fig:latticeparameter}
\end{figure}

Figure~\ref{fig:latticeparameter}a) shows the mean lattice parameters, $a$, of the initial alloy and resulting porous samples with their standard errors as determined from their peak positions
in dependence of the Williamson-Hall coherently scattering domain size $D_{\rm W-H}$
as evaluated out of the FWHM and their dependency on $q$. Theoretical $a$ values for bulk Au ($\unit[407.82]{pm}$), bulk Ag ($\unit[408.53]{pm}$) and the master alloy ($\unit[408.35]{pm}$) are indicated for comparison. Regarding the data of the alloy (note that data points for cold-worked and recovered master alloy samples superimpose on the scale of the figure), we find a negligible deviation to the theoretical $a$.

Values for $a$ found for the nanoporous samples were compared with the bulk Au value. In principle, the residual silver increases the lattice parameter value, yet the silver fraction, between $\unit[1.3]{at.\%}$ and $\unit[6.4]{at.\%}$, is too small to affect the experiment. Focusing on the slowly dealloyed samples -- that is, on samples with clean surfaces -- we find the lattice parameter to decrease with decreasing $D_{\rm W-H}$. At the smallest ligament size, the crystal lattice is compressed by $\unit[0.03]{\%}$. Lattice parameters of the ``fast dealloyed'' samples  -- that is, samples with a trend for oxygen-covered surfaces -- show a large scatter and an upward deviation from this trend.
Figure~\ref{fig:latticeparameter}b) displays the results for the slowly dealloyed (clean surface) samples in isolation. The lattice parameter is here plotted versus the inverse of the ``true'' ligament size $L_{\rm SEM}$. The straight line of best fit (dashed line) confirms the trend of compression at small size.

\subsection{Ligament size, Apparent Grain Size, and Microstrain}

\begin{figure}[tb]
\begin{center}
\includegraphics[width=0.75\columnwidth]{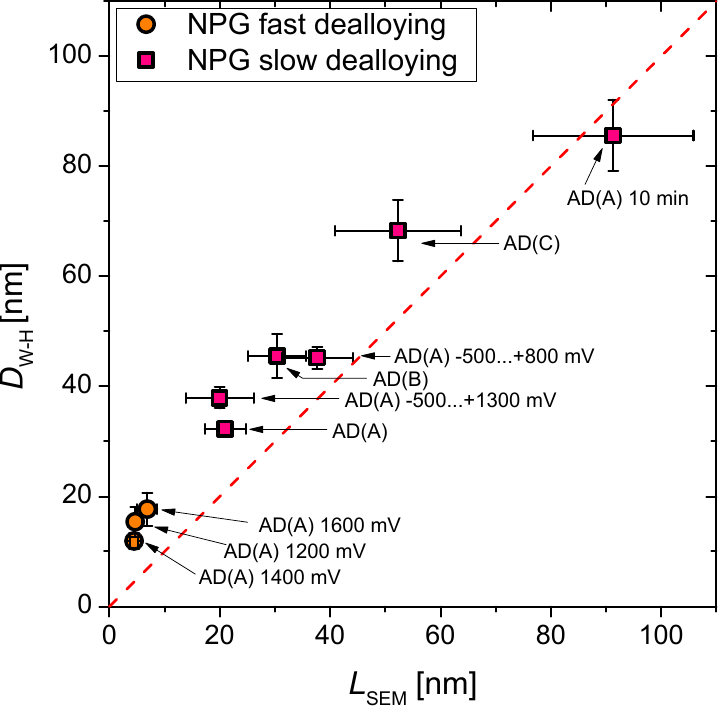}
\end{center}
\caption{Dependence of Williamson-Hall parameter, $D_{\rm W-H}$, on ligament size as determined from SEM images of nanoporous gold (NPG), $L_{\rm SEM}$, for different dealloying potentials/routes combined with subsequent electrochemical and thermal treatments (see text for sample assignment). Red dashed line indicates the direct proportionality $D_{\rm W-H} \simeq L_{\rm SEM}$.
}
\label{fig:ligamentsize}
\end{figure}

Next, we inspected whether the Williamson-Hall analysis achieves a valid separation of size and strain induced broadening. More specifically, we are interested where the ``Williamson-Hall size'', $D_{\rm W-H}$, agrees with the ``true'' ligament size as determined by SEM. By inspection of figure~\ref{fig:ligamentsize} it is obvious that the two measures for the size are indeed highly correlated. Except for the largest ligament size (100 nm), which approaches the resolution limit of the diffractometer, the Williamson-Hall size appears to agree with the true ligament size except for a constant upward shift by in the order of 15 nm.
The correlation between the true and Williamson-Hall ligament sizes is remarkable in view of the failure of our theory -- in Section \ref{theory} -- to predict a link between a simple peak-width parameter and $L$.

\begin{figure}[tb]
\begin{center}
\includegraphics[width=1.0\columnwidth]{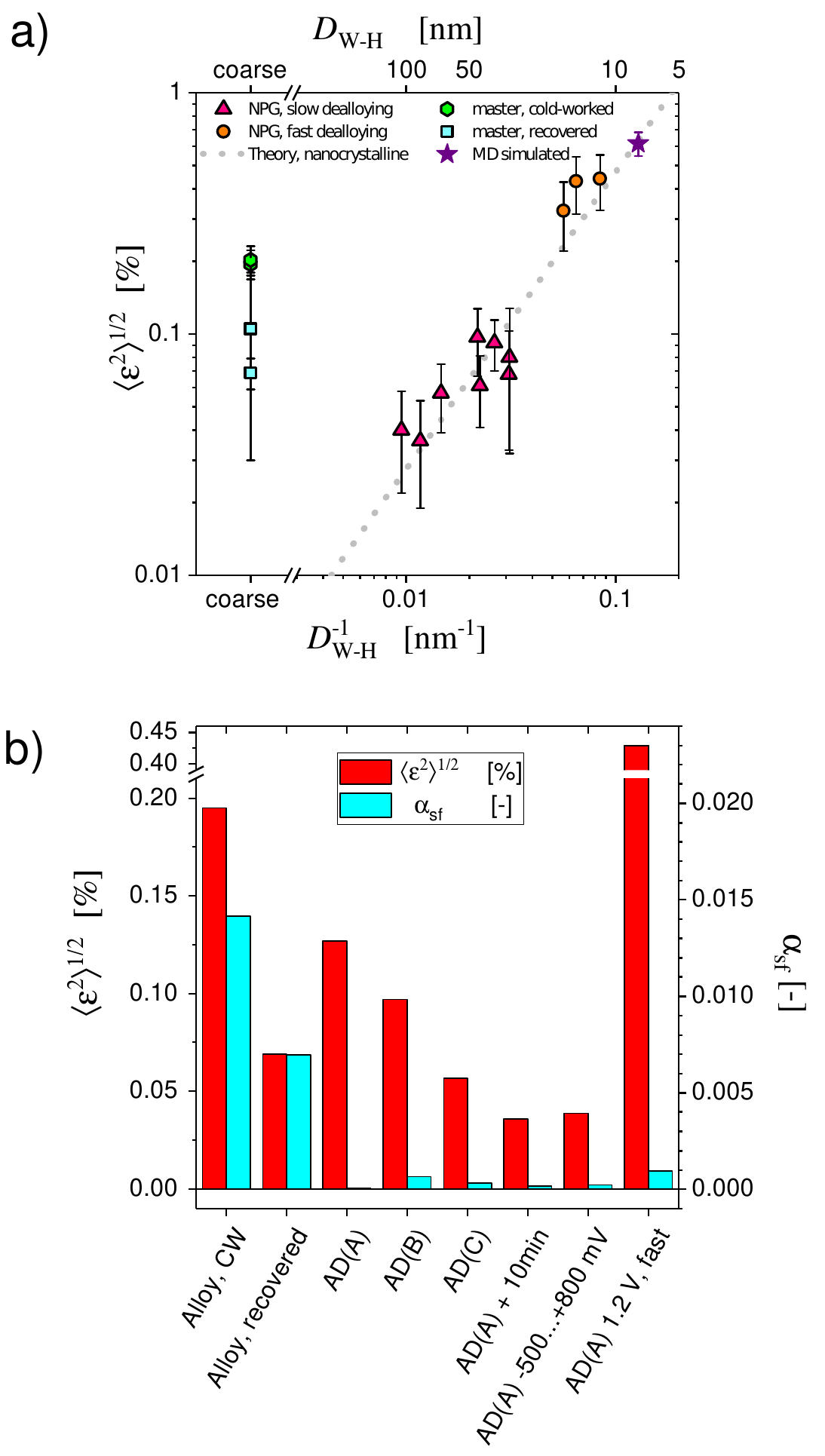}
\end{center}
\caption{Microstrain (determined from peak broadening) depending on {\bf a)} ligament size ($L_{\rm SEM}$) for Ag$_{75}$Au$_{25}$ alloys before and after thermal recovery and after different dealloying routes. Error bars correspond to the evaluated standard errors from the straight line fits (Modified Williamson-Hall). {\bf b)} Comparison of the influence on microstrain and stacking fault density $\alpha_{\rm sf}$ by different dealloying protocols and subsequent electrochemical or thermal treatments (see text for sample assignment).}
\label{fig:microstrain}
\end{figure}

The Williamson-Hall analysis separates the contributions of size and microstrain to the peak broadening. Figure~\ref{fig:microstrain}a) displays the results of the analysis for the microstrain, plotted versus  $D_{\rm W-H}$. It is seen that the results for the master alloy vary substantially, depending on whether the alloy is in its cold-worked state or recovered. The amount of variation affords an estimate of the contribution of lattice dislocations (from the cold working) to the microstrain. The microstrain values of the nanoporous samples cover a wider range, including states of very low microstrain, less than the recovered master alloy, and --especially for the fast dealloying samples-- states of very high microstrain, higher than the cold-worked master alloys. It is also observed that the microstrain values of the nanoporous samples systematically increase with decreasing ligament size. This is further emphasised by comparing the present data to the trend line from earlier work\cite{WeissmuellerRiso2001,Markmann2008} on nano\emph{crystalline} FCC metals (dashed line in the figure). The data for the size-dependence of the microstrain of the nanostructured samples is in excellent agreement with that completely independent data. The figure also shows the results of the Williamson-Hall analysis of our virtual diffraction experiment. This data is also precisely consistent with the trend.

On top of the microstrain data from the Williamson Hall analysis, we have also obtained estimates for the stacking fault density $\alpha_{\mathrm{sf}}$. This parameter, which represents the probability of a stacking fault in (111) direction, manifests itself in relative shifts in the Bragg reflection positions according to \cite{Warren1990} and is obtained along with our lattice parameter refinement, see reference~\cite{Markmann2010} for details. The light blue columns in figure~\ref{fig:microstrain}B) show the results. The most obvious finding is that $\alpha_{\mathrm{sf}}$ in the nanoporous material is systematically much lower than in the master alloy. This applies even when the nanoporous material is compared to the recovered master alloy.

The observation on the stacking faults confirms and emphasises the observation based on the microstrain data: Even though dealloying involves a massive rearrangement of the constituents of the alloy at low homologous temperature --a situation that might typically be expected to lead to defect accumulation-- the crystal lattice in the nanoporous product phase is typically more perfect than that of the initial alloy.
In order to emphasise this point we have deliberately omitted the recovery step and compared the microstrain in the cold worked master alloy to that of the nanoporous material obtained by dealloying that master alloy. Dealloying was found to decrease the microstrain from initially $(0.20\pm0.03)\%$ (cold-worked master alloy) to $(0.08\pm0.05)\%$. This is essentially the same as after a five-minute, $650^{\circ}\rm C$ anneal of the master alloy, which gave the microstrain value $(0.09\pm0.02)\%$.

\section{Discussion}

Our theory for the powder diffraction of nanoporous crystals suggests that each Bragg reflection should have two distinct components, a narrow central component that represents the long-range coherency of the crystal lattice plus a broad foot that arises from the partial loss of atomic neighbours due to the nanopores. Our virtual diffraction experiment verified the prediction by inspecting a 61-nm size single crystalline box from which atoms were removed to create the pore space. Indeed, the virtual diffraction pattern of the unrelaxed nanoporous body confirmed the theory. Yet, thermal relaxation by MD-simulation lead to a complete loss of the narrow component. Williamson-Hall analysis of the peak broadening revealed large microstrain, suggesting elastic distortion of the crystal lattice is the origin of the loss of the metal component. The elastic strain fields of lattice dislocations might be considered as an explanation. Yet the microstrain value of 0.61\% is unusually high. Furthermore, it is known that the dislocation cores tend to move into the pore space of nanoporous gold, reducing the amount of microstrain in the crystal lattice \cite{Jin2009}. One may therefore speculate that the observed broadening relates to strain gradients at the scale of characteristic ligament dimensions.

In the experimental part of our study we have analysed powder diffraction from disc-shaped samples of NPG, exploring various synthesis conditions as investigated in the earlier study reference~\cite{Graf2017}. We found that the apparent domain size, $D_{\rm W-H}$, from Williamson-Hall analysis of our diffraction data is systematically correlated with the ligament size, $L_{\rm SEM}$, as determined by SEM images. The correlation is, roughly, $D_{\rm W-H} = L_{\rm SEM} + 15 \rm nm$.  The observation of a systematically too high apparent domain size is somewhat unexpected in view of diffraction studies on nanoparticles, where the apparent domain size is invariably reduced whenever there are stacking faults or twin boundaries. On the other hand, x-ray diffraction and electron microscopy analysis typically work with different averages over the size distribution. The volume-weighted averaging in diffraction studies supplies larger values of the mean domain size when compared to the area-weighted analysis off electron micrographs \cite{Krill1998}. Furthermore, the ligament size deduced from the SEM images represents a mean ligament {\it diameter}, while the diffraction is also sensitive to the longer coherence length along the ligament {\it axis}.

The next point of the discussion treats the lattice parameters and specifically the observation that samples which were dealloyed at moderate dealloying potential and for which we expect clean surfaces showed a systematic trend for decreasing lattice parameter with decreasing ligament size.
The amount of residual Ag in these samples ranges from 1.3 to 6.4~at\%, with more silver \cite{Graf2017,Krekeler2017} and hence larger lattice parameter at smaller ligament size, see table~\ref{silvercontents}. Magnitude and sign of this trend are not consistent with the observations. Yet, the trend for lesser lattice parameter at lower ligament size is naturally consistent with the action of the relevant capillary force, the surface stress, $f$. While reversible variations in macroscopic dimension \cite{KramerNanoLetters2004} and in lattice parameter \cite{ShaoEPL2010} of NPG due to \emph{changes} in $f$ have been well documented, the \emph{absolute value} of $f$ in NPG appears not to have been measured. We now present an analysis that affords an estimate of that quantity.

The surface stress $f$ is known to result in a pressure, $p$, in the underlying solid according to \cite{WeissCahn1997}
\begin{equation}
p=\frac{2}{3}\alpha \langle f \rangle \,.
\end{equation}
The surface area per solid volume, $\alpha$, is connected to the mean ligament size $L$ via roughly
$\alpha = \frac{4}{L}$, as stated in equation~\ref{equation:alpha_cylindre} for cylindrical ligaments. The surface-induced pressure changes the volume of ligaments with an elastic bulk modulus of $B$ according to
\begin{equation}
p  = - B \frac{\Delta V}{V_0} \approx -3 B \frac{\Delta a}{a_0} \,.
\end{equation}
Combining these two equations (as in reference~\cite{Birringer2002} for nanocrystalline materials) leads to a relation for the lattice parameter,
\begin{equation}
a = a_0 - a_0 \frac{8}{9B} \,  \langle f \rangle \frac{1}{L} \,.
\end{equation}
This equation can be compared to the regression line in figure~\ref{fig:latticeparameter}. Taking the bulk modulus for Au as \unit[171]{GPa} \cite{Smithells2004} this suggests a mean surface stress value of $\langle f \rangle = \unit[3.1]{N/m}$. Typical values for unreconstructed Au surfaces determined via \emph{ab-initio} studies as presented in \cite{Umeno2007} range between $2.0 - \unit[3.3]{N/m}$. This remarkable agreement strongly hints to surface stresses as the origin of the reduction in lattice parameter in NPG at small ligament size.

Contrary to the slowly dealloyed samples, the ones produced by fast dealloying show larger $a$ and a possible trend for increased $a$ at faster dealloying. Dealloyed NPG is known to develop silver-rich clusters in the centre of its ligaments \cite{Krekeler2017,NewmanTAp2017,Mahr2017}. Taking into account the short time for structure development, it is conceivable that ligaments in fast-dealloyed NPG have a very high silver concentration in their centre, which mainly contributes to the X-ray scattering signal while a gold rich outer shell, if thin enough, does not produce noteworthy coherent scattering signal. This would result in an increased lattice parameter.
Alternatively, the trend for larger lattice parameter in the fast dealloying samples might be linked to the higher oxygen coverage in the samples along with the trend for oxygen adsorption to make the surface stress of nanoporous gold less tensile. This is supported by in-situ dilatometry data for NPG \cite{Jin2011} and by in-situ x-ray studies, which show lattice parameter expansion during oxysorption \cite{ShaoEPL2010}.

The large microstrain at small ligament size are also of interest. In their diffraction study of NPG, Schofield et al. \cite{Schofield2008} observe an asymmetry of the Bragg reflections, with a broad component shifted to lesser diffraction angles. They discuss this component as representing regions of positive (expansion) strain. In fact, when the ligaments of NPG are modelled as idealised long cylinders, then the action of surface stress is to compress the cylinder axially and expand it in the radial direction \cite{Weissmueller2010}. The Bragg reflection asymmetry in reference~\cite{Schofield2008} is therefore naturally consistent with surface-induced strain in nanoscale structures -- the same region tends to be strained differently in different orientations. Yet, our study failed to observe a significant peak asymmetry in either, the experimental or the virtual diffraction patterns. The origin of this discrepancy is unclear. One observation is that reference~\cite{Schofield2008} reported the asymmetry only for samples investigated very shortly after dealloying; this might point towards a transient state of very freshly dealloyed NPG. Such states were not explored in the present study.

The high microstrain values in the experimental part of our study are remarkable. Perfectly consistent behaviour was obtained by virtual diffraction. Furthermore, an independent determination of the microstrain based on evaluating atomic positions in relaxed nano\emph{crystalline} materials confirmed the same observation in that case. It is therefore significant that the microstrain in nanocrystalline materials --that is, massive (not porous) polycrystals with nanoscale grain size-- follow precisely the same trend as the present data. This is seen by comparing the present data to the dashed line in figure \ref{fig:microstrain}; the comparison is on an absolute scale with no fitting parameters.
Local lattice strains within the given grain arrangement (which we referred to as microstructural constraints) were caused by the discrete nature of grain sizes in different crystallographic directions and the corresponding lattice plane spacing. Since the pore space around the ligaments is empty, that explanation cannot apply here. 

As a source of the high microstrain values we may, once again, inspect the influence of surface stress. The surface-induced strains increase reciprocal to the ligament size \cite{WeissCahn1997}. The anisotropy of the strain in the bulk which is necessary to cause X-ray peak broadening does not necessarily result from an anisotropy of that surface stress state. An aspect in the ligament shape, as e.g. a cylindrical shape, causes expansive strain along the axis of the cylinder and compressive strain perpendicular to it, simply because of the geometrical conditions even for an isotropic surface stress. Such an anisotropy of the strain state in the ligaments results in microstrain when the signal is spatially averaged (what is done during calculation of the autocorrelation function in virtual diffraction and what is the case in the experimental samples because of their polycrystallinity), quite analogous to the nano\emph{crystalline} case. Another probably contributing factor comes up when following the argumentation line presented in \cite{Markmann2010}. The anisotropy of the gold crystal lattice itself already totally suffices to cause a considerable increase of microstrain with increasing stress which, in this case, is not externally applied but internally induced by the ligament surface.

\section{Conclusions}

Imaging techniques, such as electron microscopy provide only limited access to structural parameters and are in many cases sources for misinterpretations of synthesis-property relationships and fundamental corrosion processes. In contrast, diffraction experiments provide a simple access to numerical values of ligament size, lattice parameter and microstrain which are representative for the whole sample. The good correlation between the ligament size determined by analysis of SEM pictures and coherently scattering domain sizes as determined by the analysis of X-ray peak broadening as shown in figure~\ref{fig:ligamentsize} proofs the applicability of X-ray line profile analysis as a suitable tool to investigate microstructural properties of nanoporous metals.

With respect to the named synthesis-property relationships we outline the following points as key results from our investigations:
\begin{itemize}
\item Both grain size and ligament size should contribute to the peak broadening in X-ray diffraction. For NPG fabricated by dealloying thermal relaxation annihilates the lattice coherency over the (large) grain size leaving the ligament size as the only source of domain size broadening convoluted with microstrain in the X-ray reflection peaks.
\item Ligament sizes as derived by Williamson-Hall analysis of diffraction data are consistent with these observed by SEM over a range of 5 - $\unit[90]{nm}$ with a systematic  underestimation of about \unit[15]{nm}. Nevertheless, X-ray diffraction as a non-destructive technique can be utilised to determine ligament sizes in a faster way that is less prone to subjective errors.
\item NPG with very small ligament size (below \unit[10]{nm}) shows extraordinary high values of microstrain in simulated as well as in real samples. Surface stresses acting on the ligaments are the most probable reason for that.
\item The lattice parameters in NPG can be smaller than the theoretical value of Au. A systematic dependence of this deviation of the lattice parameter on the ligament size can be observed in case of slow dealloying techniques. A mean surface stress of $\langle f \rangle = \unit[3.1]{N/m}$ in agreement with \emph{ab-initio} calculations could be evaluated.
\item Comparing values of microstrain and stacking fault densities between the cold-worked initial alloy and the final NPG leads to the conclusion that dealloying represents a strategy to relax the Au lattice even more than the thermal treatment of the parent alloy.
\end{itemize}

\begin{acknowledgments}

This work was supported by German Research Foundation (DFG) through FOR2213 ``NAGOCAT'', subproject 3. The authors gratefully acknowledge the computing time granted by the John von Neumann Institute for Computing (NIC) and provided on the supercomputer \textsc{JURECA} \cite{Krause2016} at J{\"u}lich Supercomputing Centre (JSC), project HHH34.

\end{acknowledgments}

\bibliography{XRAYNPG_Refs_v13}
\bibliographystyle{apsrev}

\end{document}